\documentclass[a4paper,fleqn,usenatbib,useAMS]{mnras}

\usepackage[T1]{fontenc}
\usepackage{ae,aecompl}

\usepackage{graphicx}	% Including figure files
\usepackage{amsmath}	% Advanced maths commands
\usepackage{amssymb}	% Extra maths symbols
\usepackage{multirow}
\usepackage{longtable}

\usepackage{graphicx}	% Including figure files
\usepackage{multicol}        % Multi-column entries in tables
\usepackage{bm}		% Bold maths symbols, including upright Greek
\usepackage{pdflscape}	% Landscape pages

\title{Constraining Properties of GRB Central Engines with X-ray flares}

\author[S.-X. Yi et al.]{
Shuang-Xi Yi,$^{1,2}$
Wei Xie,$^{3}$
Shuai-Bing Ma,$^{2}$
Wei-Hua Lei$^{2}$
and Mei Du$^{1}$
\\
$^{1}$School of Physics and Physical Engineering, Qufu Normal University, Qufu 273165, China, yisx2015@qfnu.edu.cn\\
$^{2}$School of Physics, Huazhong University of Science and Technology, Wuhan 430074, China, leiwh@hust.edu.cn\\
$^{3}$School of Physics, Guizhou Normal University, Guiyang 550001, China, xieweispring@gznu.edu.cn\\
}

\pubyear{2021}

\begin{document}
\label{firstpage}
\pagerange{\pageref{firstpage}--\pageref{lastpage}}
\maketitle

% Abstract of the paper
\begin{abstract}
X-ray flares in gamma-ray bursts (GRBs) are believed to be generated by the late activities of central engine, and thus provide an useful tool to diagnose the properties of central objects. In this paper, we work on a GRB X-ray flare sample whose bulk Lorentz factors are constrained by two different methods and the jet opening angles are determined by the jet breaks in afterglow lightcurves. Considering a hyperaccreting stellar-mass black hole (BH) as the central engine of GRBs and the Blandford \& Znajek process (BZ) as the jet production mechanism, we constrain the parameters of central engine by using the X-ray flare data. We find that the BZ mechanism is so powerful making it possible to interpret both GRB prompt emissions and bright X-ray flares. The wind parameter ($p$) and accreted mass ($M_d$) fall into reasonable ranges. Our result is also applied to GRB 170817A. The late X-ray flare in GRB 170817A, if it is true, might not be a BH origin.
\end{abstract}

% Select between one and six entries from the list of approved keywords.
% Don't make up new ones.

%%%%%%%%%%%%%%%%%%%%%%%%%%%%%%%%%%%%%%%%%%%%%%%%%%

%%%%%%%%%%%%%%%%% BODY OF PAPER %%%%%%%%%%%%%%%%%%

\section{Introduction}

Gamma-ray burst (GRB) is the most luminous electromagnetic explosion in the universe.
According to the duration of $\gamma-ray$ emissions, GRBs can be divided into
two types, i.e., short GRBs with the duration time $T_{90} < 2 \,s$ and long GRBs with the duration time $T_{90} > 2 \,s$. Long GRBs are supposed
to come from core-collapse of a massive star \citep{1993ApJ...405..273W,1999ApJ...524..262M}, while short
GRBs are believed to be the merger of two compact stars \citep{1986ApJ...308L..43P,1989Natur.340..126E,1992ApJ...395L..83N}.
Different types of GRBs have different progenitors, but in order to explain both the prompt emissions and
subsequent afterglow emissions, the central engine of GRBs needs to be proposed. The central engine of GRBs
remains a long-lasting question in GRB physics, but two main contenders have been proposed for the central engine:
a hyper-accreting stellar-mass black hole (e.g., BH, \citealp{1993ApJ...405..273W,1999ApJ...518..356P,2001ApJ...557..949N,2006ApJ...643L..87G,2007ApJ...661.1025L,2017NewAR..79....1L,2018ApJ...852...20L,2013ApJ...765..125L,
2016ApJ...833..129X,2016MNRAS.458.1921S}) or a newly formed, highly magnetized, millisecond neutron star (e.g., magnetar, \citealp{1992Natur.357..472U,1998PhRvL..81.4301D}a, b; \citealp{2001ApJ...552L..35Z}; \citealp{2001A&A...369..694S}; \citealp{2006Sci...311.1127D,2010ASPC..432...81M,2013MNRAS.430.1061R,2014ApJ...785...74L,2014MNRAS.443.1779R}). And both types of central engines could be operating in GRBs.

According to the fireball model, GRB prompt $\gamma$-ray emission is produced by the internal shock
within a relativistic ejecta \citep{2004RvMP...76.1143P,2006RPPh...69.2259M,2007ChJAA...7....1Z,2015PhR...561....1K}, while the
broadband afterglow emissions are usually suggested as the interaction of a relativistic ejecta with
the circumburst medium (\citealp{1997ApJ...476..232M,1998ApJ...497L..17S,2003MNRAS.342.1131W,2005MNRAS.363...93Z,2013ApJ...776..120Y,2020ApJ...895...94Y}; ),
including X-ray, optical/IR, and radio emissions. The Neil Gehrels Swift Observatory has greatly advanced our understanding
on GRBs (\citealp{2004ApJ...611.1005G}), thanks to its rapidly response of the X-ray Telescope (XRT; \citealp{2005SSRv..120..165B}a) on board, a large fraction of X-ray afterglows have been
detected (\citealp{2007A&A...469..379E,2009MNRAS.397.1177E}), and the canonical X-ray lightcurves are proposed, including five components:
a steep decay phase, a transition from a normal decay phase to a steeper jet break phase, X-ray flare and
shallow decay (plateau) (\citealp{2006ApJ...642..354Z,2006ApJ...642..389N}; \citealp{2007A&A...469..379E,2009MNRAS.397.1177E}). Considering their physical origin, two new and peculiar components of them are
related to central engine activities, e.g., X-ray flares and plateaus. The plateaus, having a temporal with a typical index of $\alpha \sim-0.5$  (\citealp{2007ApJ...670..565L}), generally appear at 100-1000 s since the GRB trigger, and usually followed by a normal decay
($\sim1$, named external plateau) or a very sharp decay ($>2$, even to 10, named internal plateau).
The plateau phase of X-ray afterglow is still a puzzle, some theoretical models have proposed. A widely accepted model is the energy injection from a rapidly spinning magnetar (\citealp{1992Natur.357..472U,1998A&A...333L..87D}a,b; \citealp{2001ApJ...552L..35Z,2007ApJ...665..599T,2010ApJ...722L.215D,
2014MNRAS.443.1779R,2014ApJ...785...74L,2018ApJ...863...50S}). The X-ray flares usually happen at
$10^2-10^5$ s after the prompt emission, the temporal behavior and spectral properties of flares are similar to that of prompt emissions. Therefore, X-ray flares may share a similar physical origin as the prompt emission of GRBs (\citealp{2005Sci...309.1833B}b; \citealp{2006ApJ...641.1010F,2007ApJ...671.1921F,2006ApJ...653L..81L,2006ApJ...642..389N,2006ApJ...642..354Z,
2007ApJ...671.1903C,2010MNRAS.406.2113C,2011ApJ...734L..27A,2013NatPh...9..465W,2013ApJ...763...15Q,2015ApJ...803...10T,2015ApJ...807...92Y,2016ApJS..224...20Y,
2017ApJ...844...79Y}a; \citealp{2016ApJ...831..111M}a; \citealp{2016ApJ...832..161M}b). The observations of X-ray flares could provide an important clue to understand the central engine of GRBs.

The central object (BH or magnetar) producing X-ray flares is still in debate. As suggested in \cite{2021ApJ...908..242D}, who have collected a large sample of GRBs with significant jet break features appeared in the multi-wavelength afterglows. By asumming GRB jets powered by BZ mechanism in the BH hyperaccretion system, they found that the BZ mechanism is so powerful making it possible to interpret the long GRB emissions within the reasonably accretion rates for different GRBs. Therefore, a BH system is most likely operating in most GRBs. \cite{2017JHEAp..13....1Y} compared the observational results with a positive correlation between $\Gamma_0$ (the initial Lorentz factor) and the jet power from the two types of BH central engine models, i.e.,the $\nu \bar\nu$ - annihilation mechanism (non-magnetized model) and the Blandford \& Znajek mechanism
(hereafter BZ, a strongly magnetized model), and they found that some parameters are contrived for the $\nu \bar\nu$ - annihilation mechanism,
whereas the latter (BZ) mechanism can generally account for the observations, which also can be seen in \cite{2013ApJ...765..125L}. The result is further confirmed by \cite{2017ApJ...838..143X}, and by using the empirical correlation between Lorentz factor and minimum variability timescale. If X-ray fares are indeed powered by the BZ mechanism of a hyper-accreting BH system, then what can we learn about the properties of GRB central engine from these X-ray flares data? This is the task of the paper.

This paper is organized as follows. The sample selection and data analysis are presented in Section 2, and the
BH central engine model is described in Section 3. We summarize our conclusion and discuss the implications in Section 4. A concordance cosmology with parameters $H_0 = 71$ km s
$^{-1}$ Mpc$^{-1}$, $\Omega_M=0.30$, and $\Omega_{\Lambda}=0.70$ is adopted in all part of this work.

\section{Sample selection}

As discussed in \cite{2013ApJ...765..125L}, \cite{2017ApJ...838..143X} and \cite{2017JHEAp..13....1Y}, two parameters of prompt emission, i.e., the beaming corrected $\gamma$-ray luminosity $L_\gamma$ and initial Lorentz factor $\Gamma_0$, are highly related to the properties of GRB central engines, and thus used to study the properties of central objects. X-ray flares are believed to be generated by the late activities of central engine. In order to constrain the properties of central engine by X-ray flares, we extensively search for the sample of GRB X-ray flares for which the Lorentz factors can be determined.

The significant X-ray flare should contain a relatively complete structure, including the rise, peak and decay phases. \cite{2016ApJS..224...20Y}
have analyzed significant X-ray flares from the GRBs observed by Swift from 2005 April to 2015 March. The parameters of the temporal behaviour of X-ray flares can be obtained by fitting with a smooth broken power law function,
such as, the start time ($T_{\rm start}$), the peak time ($T_{\rm peak}$), the end time ($T_{\rm end}$) and fluence ($S_{\rm F}$) of flare.
While $T_{\rm start}$ and $T_{\rm end}$ can be obtained from fitting temporal
power-law curves to the rise and decay portions for one
flare, and the points on the light curve where these power laws
intersect the power law of the underlying decay curve of the flare are
defined as $T_{\rm start}$ and $T_{\rm end}$. Therefore, the rising time $T_{\rm rise}=T_{\rm peak}-T_{\rm start}$, the decaying time $T_{\rm decay}=T_{\rm end}-T_{\rm peak}$, and the duration time $T_{\rm duration}=T_{\rm end}-T_{\rm start}$ of X-ray flare, respectively. This fitting method is very similar to the method proposed by  \cite{2005ApJ...627..324N} and \cite{2007ApJ...671.1903C,2010MNRAS.406.2113C}.
The isotropic energy of flare in the 0.3-10 keV band can be obtained by $E_{\rm X,iso}=4\pi D_{\rm L}^{2} S_{\rm F}/(1+z)$,
where $z$ is the redshift, $D_{\rm L}$ is the luminosity distance, and the fluence $S_{\rm F}$ is derived from the
integration of the corresponding fitting smooth broken power-law function from the start time to the end time of the flare
the energy range of $Swift/XRT$. The observed isotropic luminosity of X-ray flare is given by $L_{\rm X,iso}=E_{\rm X,iso}/T_{\rm duration}$.

Several methods are proposed to constrain the initial Lorentz factor
of GRBs. The most popular one is using the peak time of the early afterglow onset bump as the deceleration
time of the external forward shock, and the Lorentz factor of the deceleration time is about half of the initial
one (\citealp{1999ApJ...519L..17S,2007A&A...469L..13M,2010ApJ...725.2209L,2012ApJ...751...49L,2015ApJ...807...92Y}). Most of the models estimating the bulk Lorentz
factor during the GRB prompt emission phase are inapplicable
for X-ray flares. Two new methods are thus presented to constrain the upper and lower limits on the Lorentz factor of X-ray flares.
The lower limit on Lorentz factor is based on the rising time $T_{\rm rise}$ and decaying time $T_{\rm decay}$ of X-ray flares (\citealp{2007AdSpR..40.1208W,2015ApJ...807...92Y}),  i.e.,
\begin{equation}
{\Gamma _{\rm X}} > {\left( {\frac{{{T_{\rm decay}}}}{{{T_{\rm rise}}}}}
\right)^{\frac{1}{2}}} {\left [ {\frac{1}{{2(1 - \cos {\theta
_{\rm j}})}}} \right]^{\frac{1}{2}}} \approx \theta _{\rm j}^{ - 1}{\left(
{\frac{{{T_{\rm decay}}}}{{{T_{\rm rise}}}}} \right)^{\frac{1}{2}}},
\end{equation}
where $\theta_{j}$ is the jet half-opening angle. The upper limit on Lorentz factor is given by (\citealp{2010ApJ...724..861J,2015ApJ...807...92Y})
\begin{equation}
{\Gamma _{\rm X}} \le {\left( {\frac{{L\,{\sigma _{\rm T}}}}{{8\pi {m_{\rm p}}{c^3}{R_0}}}} \right)^{\frac{1}{4}}},
\end{equation}
where $L$ and $R_0$ are the total isotropic luminosity and the initial radius of the flare outflow, $\sigma _{\rm T}$ is the Thompson cross section, $m_p$ is the rest mass of protons and $c$ is the speed of light. Generally, the observed X-ray flare luminosity can be supposed as a fraction of the total luminosity of the outflow, and $L_{\rm X,iso}=0.1L$ is often used \citep{2006MNRAS.369..197F,2010ApJ...724..861J}. The typical size of the initial radius is $R_0=10^7$ cm \citep{2007ApJ...664L...1P}.
The half-opening angle $\theta_{\rm j}$ can be calculated by
the jet break time $T_{\rm j}$ and isotropic energy $E_{\gamma, \rm iso}$ for
a homogeneous interstellar medium (ISM) case (\citealp{1999ApJ...519L..17S,1999ApJ...525..737R,2001ApJ...562L..55F,2020ApJ...900..112Z}),
\begin{eqnarray}
\theta_{\rm j} & = & 0.076 \,\,{\rm rad}\,\,\left(\frac{T_{j}}{1\ \rm
day}\right)^{3/8}\left(\frac{1+z}{2}\right)^{-3/8}  E_{\gamma,\rm iso,53}^{-1/8}  \\ \nonumber
 & & \left(\frac{\eta_{\gamma}}{0.2}\right)^{1/8}\left(\frac{n}{1\ \rm
cm^{-3}}\right)^{1/8},
\end{eqnarray}
where the efficiency of the fireball in converting the energy
in the ejecta into $\gamma$-rays $\eta_{\gamma}=0.2$ and the number density of ISM $n=1$ cm$^{\rm-3}$ are usually adopted.

Therefore, in order to estimate the collimation-corrected X-ray luminosity $L_{\rm X}$ and Lorentz factor $\Gamma_{\rm X}$, and thus constrain the central engine properties, we focus on a sample of GRBs with both X-ray flares and jet break features in the afterglow light curves. We extensively search for the candidates, our selected sample consists of 29 GRBs with X-ray flares, redshift, and jet break time. All of the X-ray flares are taken from the Swift/XRT website. Some of them have multiple flares in one GRB, and the total number of X-ray flares of our selected sample is 60. The start time ($T_{\rm start}$) of flares are taken from \cite{2007ApJ...671.1921F}, \cite{2010MNRAS.406.2113C}, \cite{2011A&A...526A..27B} and \cite{2016ApJS..224...20Y}. The jet half-opening angles ($\theta_{\rm j}$), and the lower or upper limit Lorentz factor are taken form \cite{2015ApJ...807...92Y} and \cite{2017RAA....17...53X}, respectively. All of GRBs listed in Table 1 are long GRBs. The duration time and observed isotropic luminosity of X-ray flare can be estimated by $T_{\rm duration}=T_{\rm rise}+T_{\rm decay}$ and $L_{\rm X,iso}=E_{\rm X,iso}/T_{\rm duration}$. The results are reported in Table 1.

Figure 1 shows the histogram distributions of the selected X-ray flare parameters, i.e., the start time $T_{\rm start}$, the duration $T_{\rm duration}$, the observed isotropic luminosity $L_{\rm X,iso}$ and jet opening angle $\theta_{\rm j}$. For most GRBs of our selected sample, the X-ray flares start several hundred seconds after the trigger of prompt emission (see the top left panel of Figure 1), occurring at early times and lasting around hundred seconds (see the top right panel of Figure 1). The distribution of observed isotropic luminosity of X-ray flares $L_{\rm X,iso}$ peaks at $\sim $ few times of $10^{49} \ \rm ergs\ s^{-1}$ (see the bottom left panel of Figure 1). Most GRBs have an opening angle $\theta_{\rm j}$ of $\sim 3^\circ$ (see the bottom right panel of Figure 1).

For GRBs with jet opening angle, the beaming corrected luminosity of X-ray flares $L_{\rm X}$ is given by $L_{\rm X} = f_{\rm b} L_{\rm X, iso}$, where $f_{\rm b} = 1-\cos \theta_{\rm j}$ is the beaming factor. We plot $\Gamma_{\rm X}$ versus $L_{\rm X}$ (the top and bottom of the vertical lines represent for the upper and lower limits of $\Gamma_{\rm X}$ respectively) in Figure 2. For comparison, we also plot $\Gamma_{\rm 0}$ versus $L_{\gamma}$ for prompt emissions (blue dots in Figure 2, reproduced with the data from \citealp{2017JHEAp..13....1Y}). The blue solid line shows the best-fit to the prompt emission data, i.e., $\log \Gamma_0=(-4.77\pm 4.06)+(0.14\pm 0.08) \log L_{\rm \gamma}$ (see \citealp{2017JHEAp..13....1Y}). The model prediction from the BZ model is shown with blue dashed line.  We extend these results (blue solid and dashed lines) to the X-ray flares regime. As we can see from Figure 2,  there is a potential but weak correlation of $\Gamma_{\rm X} - L_{\rm X}$ in X-ray flares data. We also find that the blue lines just lay between the upper and lower limit of Lorentz factors for most X-ray flares. These results indicate that the central engine for X-ray flares and prompt emissions can be the same.

%We extend these results (blue solid and dashed lines) to the X-ray flares regime. As we can see from Figure 2, the X-ray flares data also exhibit a correlation of $\Gamma_{\rm X} - L_{\rm X}$, which is consistent with the model predictions and the extension of the best-fit to the prompt emission. Our result thus suggest that X-ray flares might share a similar physical origin with the prompt emissions.

%\clearpage
\begin{figure*}
\includegraphics[angle=0,scale=0.30]{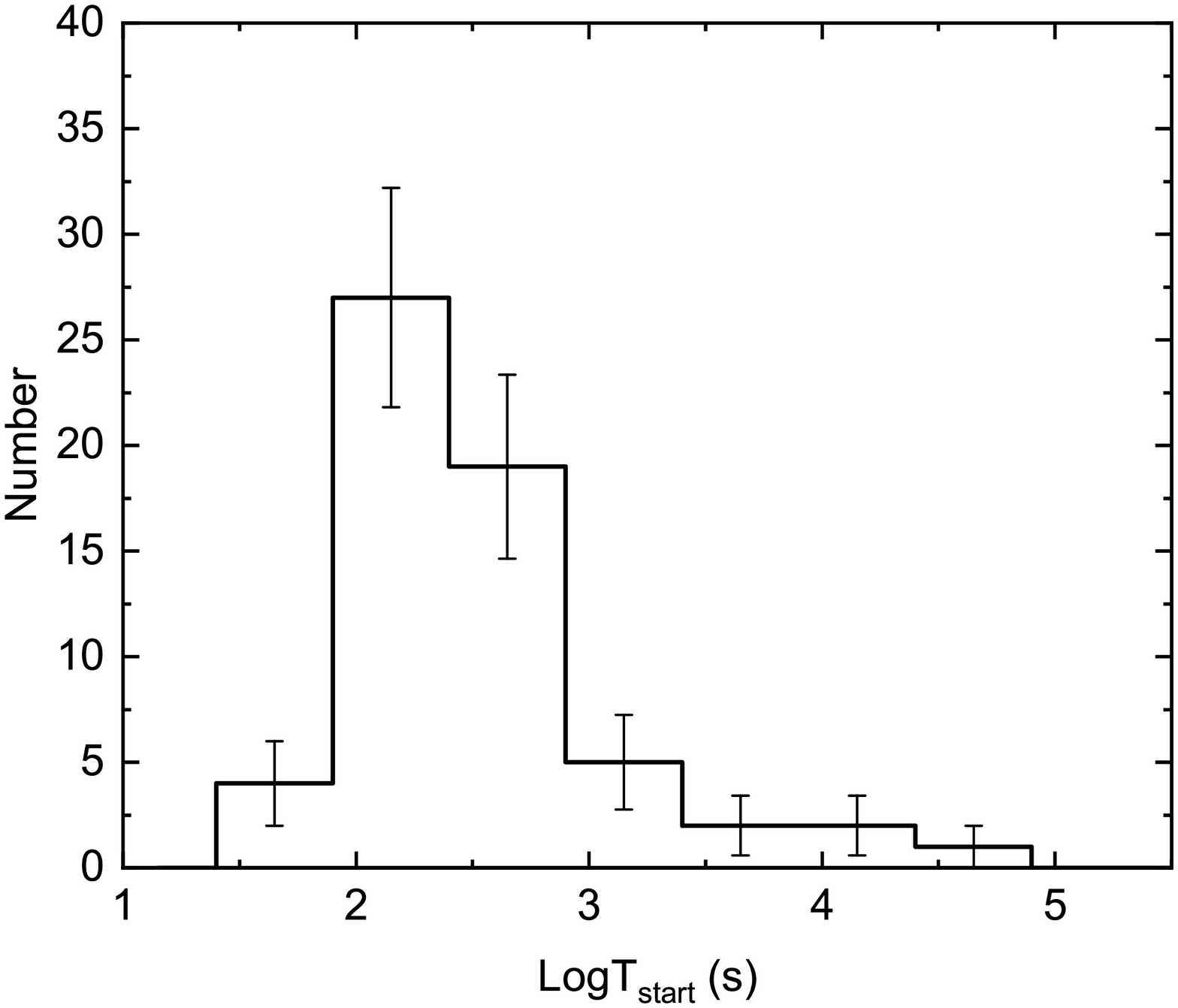}
\includegraphics[angle=0,scale=0.30]{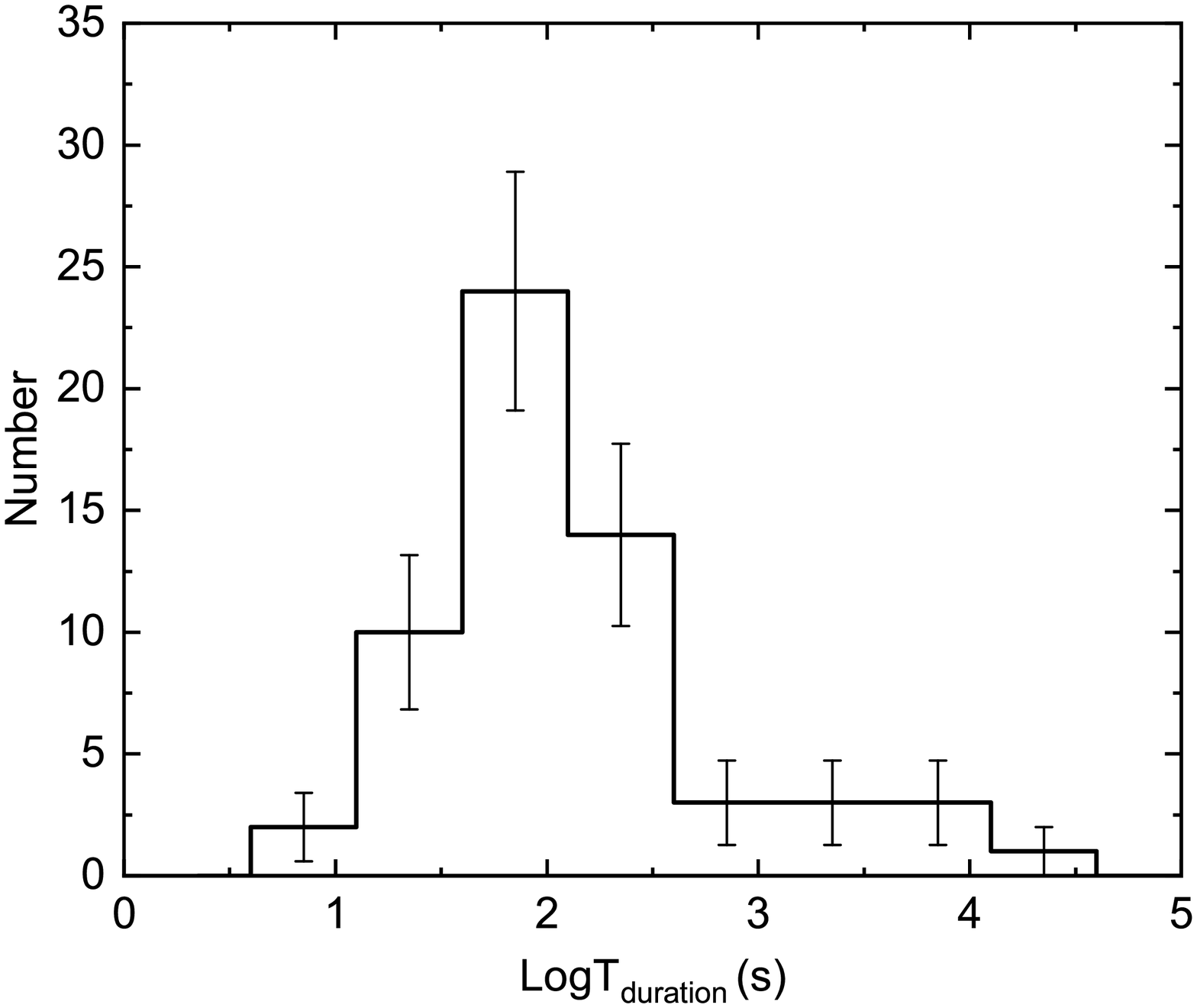}
\includegraphics[angle=0,scale=0.30]{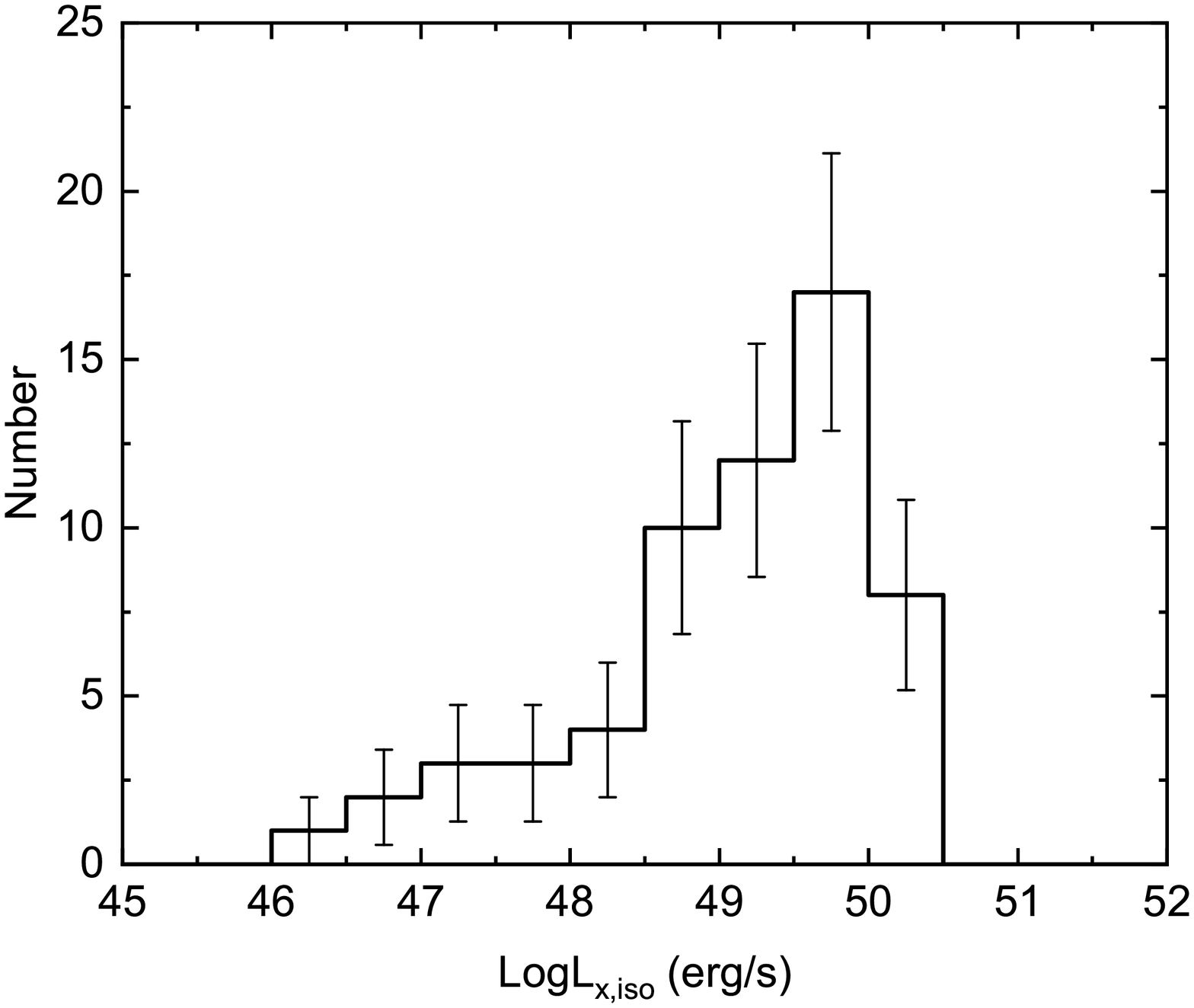}
\includegraphics[angle=0,scale=0.30]{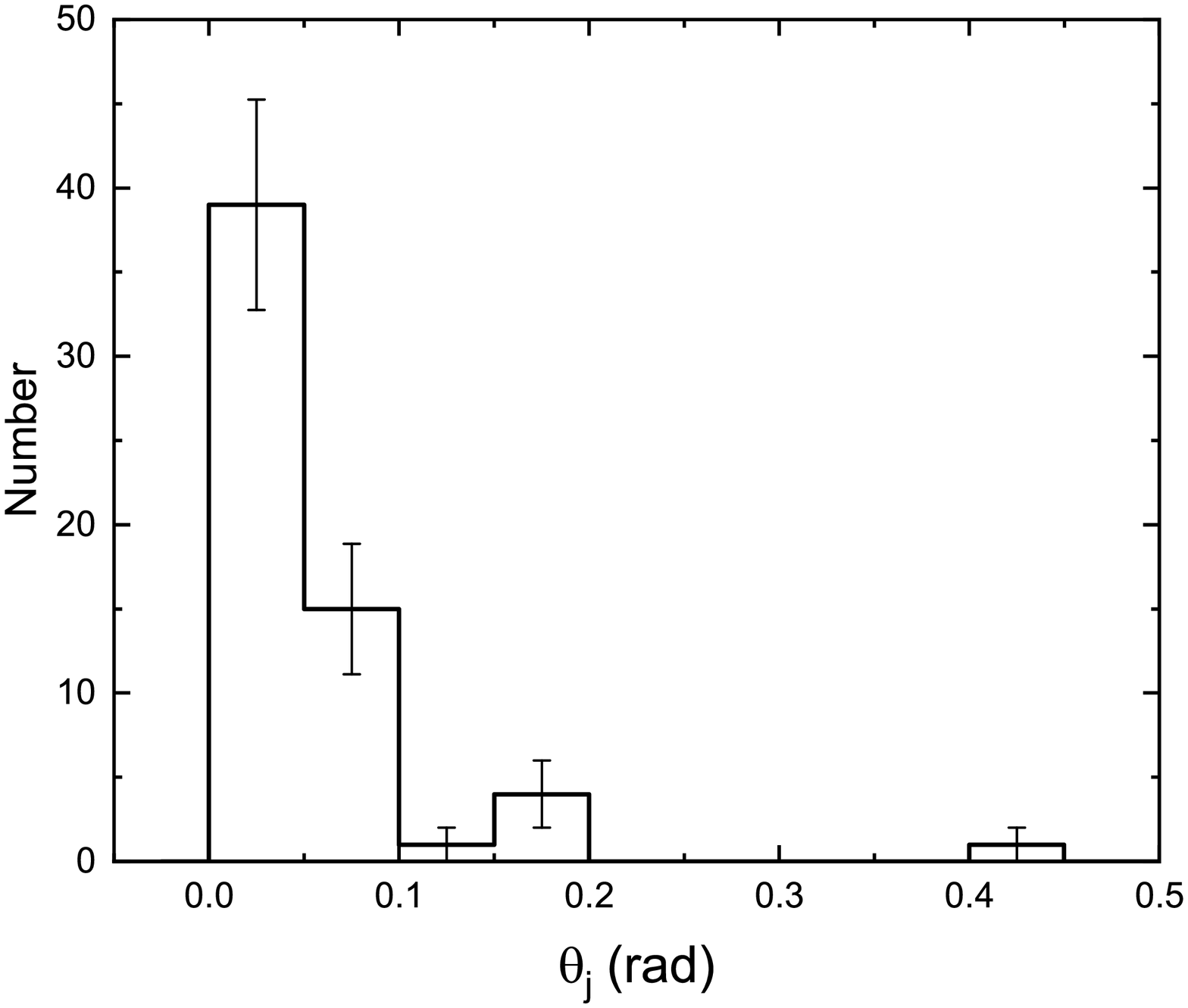}
\caption{The distributions of parameters of selected GRB X-ray flares, i.e., the start time $T_{\rm start}$ (top left), the duration $T_{\rm duration}$ (top right), X-ray isotropic luminosity $L_{\rm X,iso}$ (bottom left) in 0.3-10 keV band and jet opening angle $\theta_{\rm j}$ (bottom right). Our X-ray flare sample as described in Section 2.}
\end{figure*}

\begin{figure*}
\center
\includegraphics[angle=0,scale=0.30]{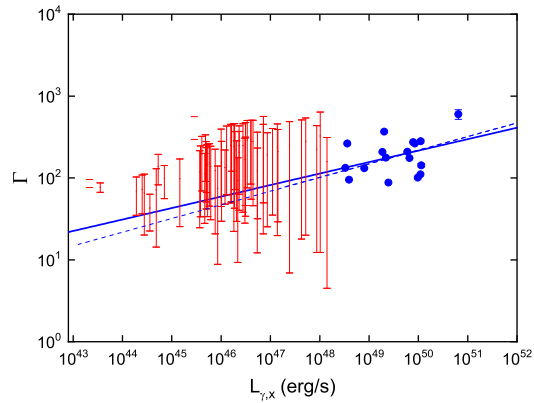}
\caption{Lorentz factor $\Gamma$ vs. beaming corrected luminosity of X-ray flares ($L_{\rm X}$, the red vertical lines) and prompt GRBs ($L_{\gamma}$, blue dots). The top and bottom of the vertical lines represent the upper and lower limit of Lorentz factor for X-ray flares. The blue lines are adopted from \citep{2017JHEAp..13....1Y}. The blue solid line shows the best-fit to the prompt emission data $\Gamma_0 \propto L_\gamma^{0.14}$, and the blue dashed line corresponds to the model prediction from the BZ model (see \citealp{2017JHEAp..13....1Y}). }
\end{figure*}

\section{The BH central engine model}
Two popular models have been proposed for GRB central engines, i.e., a hyper-accreting stellar-mass black hole  and a newly formed millisecond magnetar.  For each GRB, the type of central object (BH or magnetar) is still unclear. Recently, the comprehensive studies by \cite{2021ApJ...908..242D} with the \textit{Swift} data suggested that a BH is most likely operating in most GRBs. In BH central engine model, there are two main energy reservoirs to power the jet: the accretion energy in the disk that is carried by neutrinos and anti-neutrinos, which annihilate and power a bipolar non-magnetized outflow (\citealp{1993ApJ...405..273W,1999ApJ...518..356P,2001ApJ...557..949N,2007ApJ...661.1025L,2017NewAR..79....1L,2009ApJ...700.1970L,
2013ApJ...765..125L,2017ApJ...849...47L,2016ApJ...833..129X,2016MNRAS.458.1921S}); and the BH spin energy, which can be tapped by a magnetic field connecting the remote astrophysical load through the BZ mechanism  (\citealp{1977MNRAS.179..433B,2000PhR...325...83L,2000PhRvD..61h4016L,2002MNRAS.335..655W,2005ApJ...630L...5M,2005ApJ...619..420L,
2013ApJ...765..125L,2017ApJ...849...47L,2011ApJ...740L..27L,2021ApJ...908..242D,Chen2021}). Usually, a BZ jet is more powerful. During the X-ray flare stage, the typical accretion rate is far below the igniting accretion rate, the neutrino-annihilation power cannot account for the luminous X-ray emissions. Therefore, the BZ model is likely responsible for the X-ray flares (\citealp{2017ApJ...849...47L}.) More details please see the recent reviews about BH central engine model (i.e., \citealp{2017NewAR..79....1L,2019ChA&A..43..143L}).

\cite{2017JHEAp..13....1Y} extended the empirical correlations $\Gamma_0 -  E_{\gamma,iso}$ (by \citealp{2010ApJ...725.2209L}) and $\Gamma_0 -  L_{\gamma,iso}$ (by \cite{2012ApJ...751...49L}) to the correlations between $\Gamma_0$ and the beaming corrected $\gamma$-ray energy $E_\gamma$ and luminosity $L_\gamma$, and compared them with the predictions from these two types of jet production mechanism in the context of BH central engine model. Their results favor the BZ mechanism, which has been further confirmed by \cite{2017ApJ...838..143X} with the empirical correlation between Lorentz factor and minimum variability timescale.

Inspired by these findings, we dedicate to investigate the properties of black-hole (BH) central engine with the X-ray flares sample obtained in Section 2. The BZ power from a BH as a function of its accretion rate $\dot{M}_{\rm in}$ and spin $a_\bullet$ is (\citealp{1977MNRAS.179..433B,2000PhR...325...83L,2000PhRvD..61h4016L,2002MNRAS.335..655W,2005ApJ...630L...5M,2008ChJAA...8..404L,
2013ApJ...765..125L,2011ApJ...740L..27L,2021ApJ...908..242D})

\begin{equation}
\dot{E}_\text{B}=9.3\times 10^{53} a_\bullet^2 \dot{m}_{\rm in} X(a_\bullet) \rm{\ erg\ s^{-1}},
\label{eq:BZ}
\end{equation}
and
\begin{equation}
X(a_\bullet)=F(a_\bullet)/\left(1+\sqrt{1-a_\bullet^2}\right),
\end{equation}
\begin{equation}
F(a_\bullet)=[(1+q^2)/q^2][(q+1/q)\arctan{q}-1].
\end{equation}
Here $q=a_\bullet/(1+\sqrt{1-a_\bullet^2})$. $\dot{m}_{\rm in}\equiv\dot{M}_{\rm in}/M_\odot \rm s^{-1}$ denotes the accretion rate normalized with solar mass per second at the inner edge of the disk.

The baryon loading is essential in determining the Lorentz factor of GRB jet. As argued in \cite{2013ApJ...765..125L}, the accretion disk is the major source of the baryon loading of GRB jet. As to the X-ray fare phase we interested, the accretion rate is not that high enough to trigger efficient neutrino emission and the hyperaccreting disk in principle enters the regime of advection-dominated accretion flow (ADAF). A typical feature of ADAF is strong disk wind due to the positive Bernoulli parameter (see also \citealp{2018MNRAS.477.2173S} and references therein). Especially, \cite{2015ApJ...799..71G} argued that the wind will always occur in the accretion flows as long as the cooling rate cannot balance the viscous heating rate, regardless of that the flow is optically thin or thick. In this work, it is such a disk wind that drives the baryons into the jet. Due to the existence of the wind, the accretion rate will change with the disk radius, which could be described as follow for simplicity,
\begin{equation}
\dot{M}=\dot{M}_\text{out}\left(r/r_\text{out}\right)^p,
\label{eq:dotM}
\end{equation}
where $p$ is a scalar parameter. The Equation (\ref{eq:dotM}) was firstly introduced by \cite{1999MNRAS...310...1002S} as a conclusion of their hydrodynamic (HD) simulation of a non-radiative accretion flow. The subsequent radiation-hydrodynamical (RHD) simulations (e.g., \citealt{2005ApJ...628...368O}) and magnetohydrodynamical (MHD) simulations (e.g., \citealt{2001MNRAS...322...461S, 2002ApJ...573...738H, 2003ApJ...592...1042I, 2003ApJL...596L...207P}; see review in \citealt{2012ApJ...761...129Y}) obtain similar results that typically $p \sim 0.5-1$. Although \cite{1999MNRAS...310...1002S} ascribed the physical origin of $p$ to the vigorous convection in the flow, some later researches (e.g., \citealt{2012MNRAS...426...3241N, 2012ApJ...761...130Y} and \citealt{2015...804...101Y}) argued that the accretion flow is convectively stable if there are magnetic-fields in the flow, i.e., the inward decrease of accretion rate should be caused by the disk outflow/wind. Considering the existence of the magnetic field in our model, we interpret the parameter $p$ as the disk wind indicator. \cite{2016ApJ...832..161M}b tested the central engine models by using several GRBs with X-ray flares at extremely late time. They fixed the value of wind parameter $p \sim1$. As a result, the mass accretion rate onto BH horizon will be extremely low, especially for those X-ray flares at very late time. And they concluded that BZ mechanism is not strong enough to power these late time X-ray flares. However,  the wind physics is poorly understood, and the value of $p$ is quite uncertain. Here, we take $p$ as a free parameter, and then constrain its value with our sample of X-ray flares. The mass loss rate through wind of the disk from $r$ to $r+dr$ is
\begin{equation}
\dot{m}_\text{w}=\frac{1}{4\pi r}\frac{d\dot{M}}{dr}.
\label{eq:dotmw}
\end{equation}
Giving the half open angle $\theta_{\rm j}$ of the jet, the baryon loading from the disk from $r$ to $r+dr$ can be expressed as
\begin{equation}
d\dot{M}_{\rm j}=(1-\cos\theta_{\rm j})\dot{m}_\text{w}4\pi r dr.
\label{eq:ddotMj}
\end{equation}
The total mass loading into the jet is
\begin{equation}
\dot{M}_{\rm j}=\int_{r_{\rm in}}^{r_{\rm out}} d \dot{M}_{\rm j}.
\label{eq:dotMj}
\end{equation}
Integrating the equations (\ref{eq:dotM})-(\ref{eq:dotMj}) gives
\begin{equation}
\dot{M}_{\rm j}=(1-\cos\theta_{\rm j})(\dot{M}_\text{out}-\dot{M}_\text{in})\\
=(1-\cos\theta_{\rm j})\dot{M}_\text{out}\left[1-\left(\frac{r_{\rm in}}{r_{\rm out}}\right)^p\right].
\label{eq:dotMjet}
\end{equation}
Ignoring the details of converting the Poynting energy from the BH to the bulk kinetic energy of the jet, we can estimate the Lorentz factor of the jet as,
\begin{equation}
	\Gamma \approx \frac{\dot{E}_{\rm B}}{\dot{M}_{\rm j}c^2}=0.52a_\bullet^2(1-\cos\theta_{\rm j})^{-1} X(a_\bullet)\xi_{\rm out}^{-p}(1-\xi_{\rm out}^{-p}),
\label{eq:GammaBZ}	
\end{equation}
here $\xi_{\rm out}$ denotes the ratio of the outer radius to the inner radius of the disk, i.e., $\xi_{\rm out} \equiv r_{\rm out}/r_{\rm in}$.  The innermost radius $r_{\rm in}$ is defined as
\begin{equation}
r_{\rm in}=r_{\rm g} \left\{3+Z_2 - [(3-Z_1)(3+Z_1+2Z_2)]^{1/2}\right\},
\end{equation}
for $0\leq a_{\bullet} \leq 1$, where $r_{\rm g} =G M_\bullet /c^2$, $Z_1 \equiv 1+(1-a_\bullet^2)^{1/3} [(1+a_\bullet)^{1/3}+(1-a_\bullet)^{1/3}]$,  and $ Z_2\equiv (3a_\bullet^2+Z_1^2)^{1/2}$.

Given the parameters such as $a_\bullet$, $\xi_{\rm out}$ and $\theta_{\rm j}$, we can infer the disk wind indicator parameter $p$ by comparing the theoretical expression (Equation (\ref{eq:GammaBZ})) for Lorentz factor $\Gamma$ to the observational data (see Table 1).

In addition, by introducing a radiation efficiency $\eta$, we can relate the X-ray flare luminosity to the BZ power (see Equation (\ref{eq:BZ}) ) as,
\begin{equation}
L_{\rm X,iso}(1-\cos\theta_{\rm j})=\eta \dot{E}_\text{B}.
\label{eq:Lbz}
\end{equation}
Consequently, substituting Equation (\ref{eq:Lbz}) into Equations (\ref{eq:BZ}) and (\ref{eq:dotM}), one can figure out the accretion rate at the outer edge of the disk $\dot{M}_{\rm out}$. Finally, one can get the total mass of the accreted matter by multiplying with the flare duration, i.e.,
\begin{equation}\label{key}
M_{\rm d}=\dot{M}_{\rm out} T_{\rm duration}/(1+z).
\end{equation}

Therefore, using the rich X-ray flare data, we can investigate the circular environment (total mass supply) and accretion physics (type of the disk and strength of  the wind) of BH central engine.

%&&&&&&&&&&&&&&&&&&&&&&&&&&&&&&&&&&&&&&&&&&&&&&&&&&&&&&&&&&&&&&&&&&&&&&&&&&&&&&&&&&&&&&&&&&&&&&&&&&&&&&&&&&

\section{Results on the disk wind and the total accreted mass associated with the X-ray flare}

\begin{figure*}\label{fig:3}
\includegraphics[angle=0,scale=0.30]{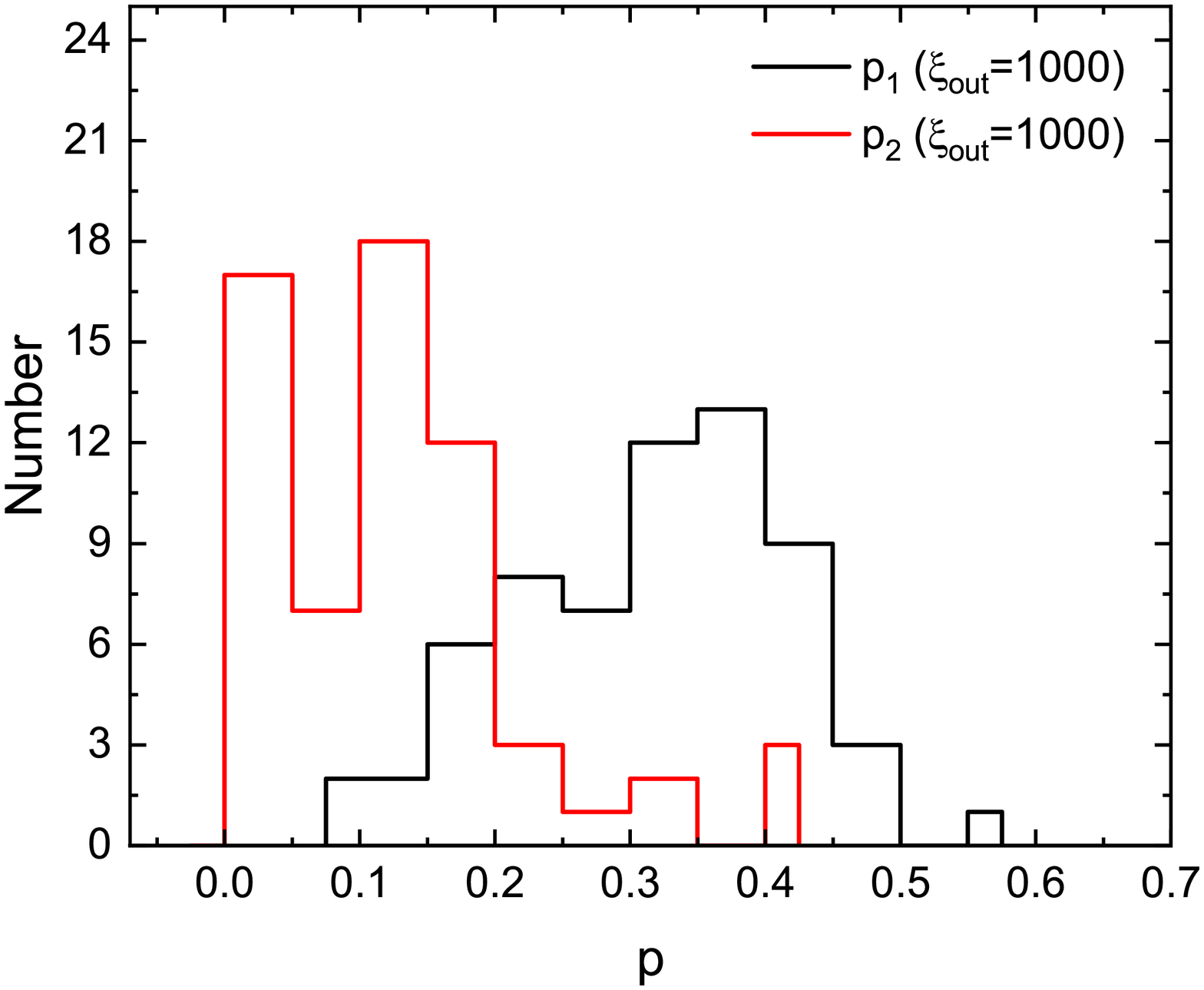}
\includegraphics[angle=0,scale=0.30]{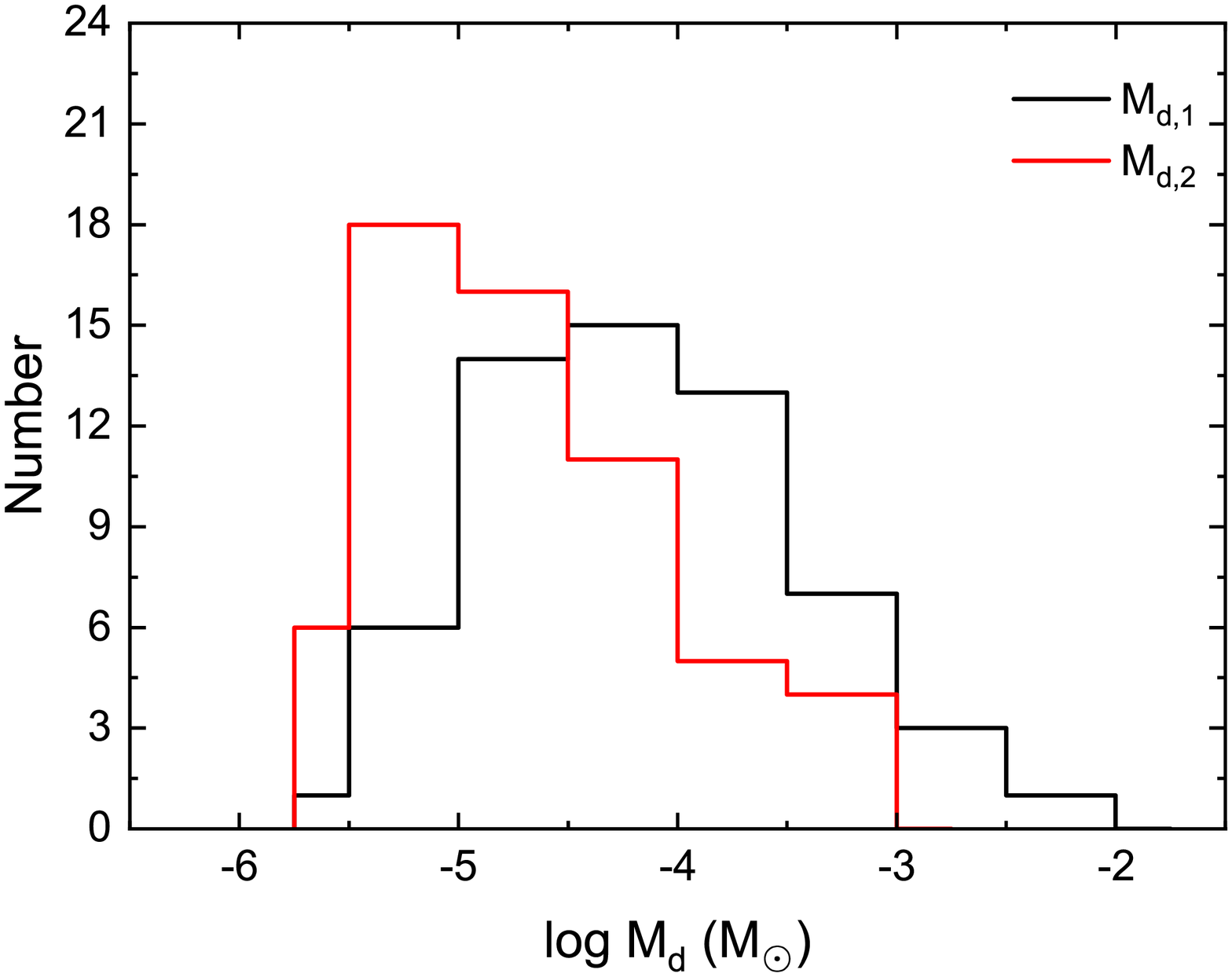}
\caption{The distributions of parameters $p$ and $M_{d}$ of GRB X-ray flares. We study two cases according to the two estimations of Lorentz factors, i.e., $p_1$, $M_{d,1}$ for $\Gamma_{\rm X}=\Gamma_{\rm lower}$ (Case 1, with black lines), and $p_2$,  $M_{d,2}$ for $\Gamma_{\rm X}= \Gamma_{\rm upper}$ (Case 2, with red lines).}
\end{figure*}

\begin{figure*}\label{fig:4}
\includegraphics[angle=0,scale=0.30]{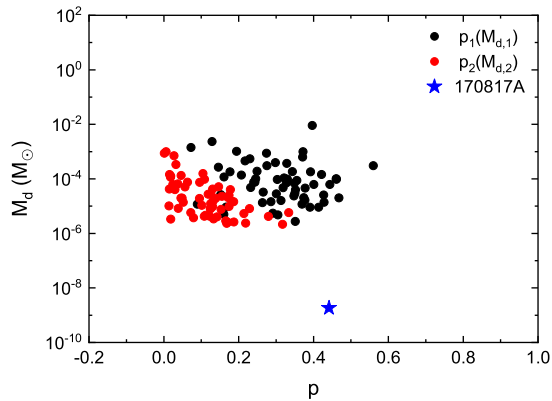}
\caption{The $p$ versus $M_{d}$ for our GRB X-ray flare sample (red and black dots have the same meanings as Figure 3) and GRB 170817A (blue star). GRB 170817A is a significant outlier, implying a distinct central engine with other GRBs.}
\end{figure*}

In our model, we consider a fast spinning BH with a typical value of the spin parameter $a_{\bullet}=0.95$, and we assume $\xi_{\rm out}=1000$ for the hyperaccreting disk. Due to the lack of the knowledge about the concrete dissipation process of the jet, we take $\eta=0.1$ as a typical value, like done by some previous works (such as \citealt{2016ApJ...832..161M}b, \citealt{2017ApJ...849...119C}). For each GRB X-ray flare, both the lower limit ($\Gamma_{\rm lower}$) and upper limit ($\Gamma_{\rm upper}$) of the Lorentz factor are calculated (as shown in Table 1). We thus have two sets of wind parameter $p$ and disk mass $M_{\rm d}$, i.e., $p_1$, $M_{d,1}$ for $\Gamma_{\rm X}=\Gamma_{\rm lower}$ (Case 1), and $p_2$,  $M_{d,2}$ for $\Gamma_{\rm X}= \Gamma_{\rm upper}$ (Case 2). The results are listed in Table 1. The histogram distributions of the parameters are shown with different colors for these two cases (black color for Case 1 and red for Case 2) in Figure 3. One finds that $p_1>p_2$ and $M_{\rm d,1}>M_{\rm d,2}$ for each X-ray flare. This is reasonable, since a strong disk wind (large value of $p$) leads to a heavy baryon loading to the jet and thus a small Lorentz factor. At same time, strong wind needs more mass supply $M_{\rm d}$ at the outer boundary.

The real values of the parameters ($p$, $M_{\rm d}$) possibly lay between ($p_1$, $M_{\rm d,1}$)  and  ($p_2$, $M_{\rm d, 2}$). Our studies hence suggest that, for most GRBs, the central engine should harbor a hyperaccreting BH with wind parameter $p$ in the range of $0.1$ to $0.6$ and the total mass supply $M_{\rm d}$ in the range of $3\times 10^{-6} M_{\bigodot} $ to $3 \times 10^{-4} M_{\bigodot} $ at late time.

Nevertheless, it's necessary to point out that the results are still far from mature due to some uncertainties in the model. In Equation (\ref{eq:GammaBZ}) we assumed all BZ power converts to the bulk motion of the jet. This assumption overestimates the ability of BZ mechanism in accelerating the jet, whereas the real wind parameter $p$ and disk mass $M_{\rm d}$ should be smaller than that above. In addition, the radiation efficiency $\eta$ in Equation (\ref{eq:Lbz}) and the disk size parameter $\xi_{\rm out}$ in Equation (\ref{eq:GammaBZ}) are not easy to be  determined in our model, both of the current value assumptions for these two parameters bring extra uncertainties, i.e., a smaller $\xi_{\rm out}$ would lead to a larger $p$, while a larger $\eta$ would lead to a smaller $M_{\rm d}$. To lower the uncertainties, further studies on the disk formation as well as the jet acceleration and dissipation are needed.

\section{Discussion}
The discovery of the association between GW 170817 and short GRB
170817A, marked the beginning of the era of multi-messenger astronomy (\citealp{2017PhRvL.119p1101A}). Short GRBs and long GRBs may share a similar central engine. However, the final product of the merger of binary NSs for GW 170817 is still a mystery. The kilonova data can be explained by either a NS central engine (\citealp{2018ApJ...861..114Y}) or a BH one (\citealp{2018ApJ...852L...5M}). The X-ray flares may also provide key information to answer this question. Especially, a suspected X-ray flare of GRB 170817A at time interval $ \sim 137-164$ days is discovered by \cite{2019MNRAS.483.1912P} and \cite{2019MNRAS.486.4479L}. They argued that this X-ray flare is an evidence of magnetar operating at $\sim 160$ days (\citealp{2019MNRAS.483.1912P,2019MNRAS.486.4479L}). To check whether this flare a result of BH activity, we constrained the wind parameter $p$ and total disk mass $M_{\rm d}$ required for this X-ray flare with the parameters ( $\Gamma \sim 10$, $\theta_{j} \sim 0.05$ rad, $L_{X,iso}=1.54\times10^{39}$ erg/s, $T_{start}\sim 137$ days, Duration $\sim 24$ days and $D_L=40.7$ Mpc) (\citealp{2019MNRAS.483.1912P,2018Natur.561..355M,2019MNRAS.486.4479L,2019ApJ...886L..17H}), and compared with that from our sample of 31 GRBs, as shown in Figure 4. The constraining results with our sample are displayed with black (Case 1: $p_1$ and  $M_{\rm d,1}$) and red dots (Case 2: $p_2$, $M_{\rm d, 2}$). The parameters ($p=0.44$ and $M_d=1.84\times10^{-9}$ $M_{\odot}$) for GRB 170817A is shown with blue star. If this X-ray flare of GRB 170817A is truly exist, and a newly formed BH indeed operates at $\sim 160$ d after the merger, the central engine should behavior in similar way with other GRBs. The blue star is thus most likely to appear somewhere in the group of red and black dots. The negative result indicate that the X-ray flare GRB 170817A might not be a BH origin. We, therefore, suggest that final product of GW 170817 is a massive NS instead of collapsing to a BH, if the X-ray flare discovered at $\sim 100$ d is present. However, the analysis of the entire sample of CXO observations of GW 170817 by \cite{2019ApJ...886L..17H} showed no statistical evidence of such a X-ray flare, and thus does not quantitatively support the magnetar model at $\sim 160$ d. Therefore, the possibility of a new-born BH as the final product of GW 170817 can not be ruled out.

We should propose the issue of achromatic breaks in the multiwavelength afterglows that are possibly `hidden' and the light curves are therefore mistaken as being single
power laws, or considered as the camouflaged achromatic breaks. Since the jet break time is very sensitive to observer angle, the jet opening angle and true energetics of GRBs
will be influenced when using the the camouflaged achromatic breaks \citep{2008MNRAS.386..859C,2013ApJ...767..141V,2018ApJ...859..160W}. Therefore, in order to obtain the jet opening angle and true energetics of GRBs, more GRBs with achromatic
breaks are needed when performing statistical analyses \citep{2020ApJ...900..112Z}. Our results also implied that the BZ mechanism as applied here is capable of producing both prompt emissions and X-ray flares. In future, a consistent study from prompt emission to X-ray flares is necessary to comprehend the evolution of central engines. Our treatment for baryon loading at low accretion rates is very primitive. The GRMHD simulations will be helpful for understanding this issue. Besides, some GRBs with plateau features have also been appeared in our selected sample, such as, GRB 060714 and GRB 060729 \citep{2007ApJ...662..443G,2007ApJ...665..554K}, the central engine for those GRBs may be a magnetar engine. However, in this work, we just focus
on the BH central engine model, and try to constrain the properties of BH by using GRB X-ray flare sample. In future, we
will study different types of GRB central engine models.

\section{Conclusion}

X-ray flares are useful tool to study the central engine of GRBs. Compared with prompt emission phase, the rich observational data as well as the long duration of X-ray flares might provide more information of central object. In this paper, we analyzed 31 GRBs with X-ray flares and estimation of Lorentz factors. Most of our selected X-ray flares are occurring at early times. We found that the $\Gamma_{\rm X} - L_{\rm X}$ correlation can be a natural extension of the $\Gamma_0 - L_{\gamma}$ proposed by \cite{2012ApJ...751...49L} and \cite{2017JHEAp..13....1Y} for prompt emissions. Therefore, X-ray flares and prompt emissions should have similar central engines. A BH is most likely operating in most GRBs (\citealp{2021ApJ...908..242D}). We constrained the properties of BH central engine by using the X-ray flares sample. For most GRBs, the wind parameter $p$ should be in the range of $0.1$ to $0.6$ and the total mass supply $M_{\rm d}$ in the range of $3\times 10^{-6} M_{\bigodot}$ to $3 \times 10^{-4} M_{\bigodot}$ for BH activities at late time. We find that the BZ mechanism is so powerful making it possible to interpret both GRB prompt emissions and bright X-ray flares.

\section*{Acknowledgements}
We thank the anonymous referee for constructive and helpful comments.
This work is supported by the National Natural Science Foundation
of China (Grant Nos. U2038106, 11703010, U1931203, U2038107), the Special Funds for Theoretical Physics of the National Natural Science Foundation of China (No. 11847102).

\section*{Data availability}
The data underlying this article are available in the article.

%%%%%%%%%%%%%%%%%%%%%%%%%%%%%%%%%%%%%%%%%%%%%%%%%%

%%%%%%%%%%%%%%%%%%%% REFERENCES %%%%%%%%%%%%%%%%%%

% The best way to enter references is to use BibTeX:

%\bibliographystyle{mnras}
%\bibliography{example} % if your bibtex file is called example.bib

% Alternatively you could enter them by hand, like this:
% This method is tedious and prone to error if you have lots of references

%%%%%%%%%%%%%%%%%%%%%%%%%%%%%%%%%%%%%%%%%%%%%%%%%%
\begin{table*}\tiny %\footnotesize
\centering
\caption{The Parameters of the GRB X-ray flares and the constrains on the properties of the central engines. }\label{}
\begin{tabular}{ccccccccccccc}
\hline
 GRB & $z$ & ${T_{\rm start}}^{a}$ & ${T_{\rm duration}}^{a}$ & ${L_{\rm X, iso}}^{b}$ & ${\theta_{\rm j}}^{c}$ & $\Gamma_{\rm lower}$ & $\Gamma_{\rm upper}$ & $p_1$ & ${M_{\rm d,1}}^{d}$ & $p_2$ & ${M_{\rm d,2}}^{d}$ & $Refs^{e}$ \\
\hline
\hline
%050416 A	&	0.65	&	1.5E6	&	501000	&	7.65E+43	&	0.026	&	43.7	&	14.6	&	0.57	&	9.01E-6	&	0.42	&	3.13E-6		\\
%050802	&	1.71	&	312	&	145	&	1.04E+48	&	0.028	&	14.9	&	157.1	&	0.55	&	2.12E-5	&	0.24	&	2.45E-6		\\
050814 (1)	&	5.3	&	1133	&	841	&	2.36E+47	&	0.046	&	36.6	&	108.5	&	0.29	&	5.52E-6	&	0.17	&	2.35E-6 &1,1,2,2		\\
050814 (2)	&	5.3	&	1633	&	944	&	2.62E+47	&	0.046	&	20.1	&	111.4	&	0.37	&	1.18E-5	&	0.17	&	2.88E-6	&1,1,2,2	\\
050820 A	&	2.62	&	200	&	182	&	6.07E+49	&	0.169	&	12.3	&	434.5	&	0.13	&	0.00234	&	0.01	&	0.001 &1,1,2,2		\\
050904 (1)	&	6.29	&	343	&	227	&	7.06E+49	&	0.045	&	20.8	&	451.3	&	0.37	&	6.36E-4	&	0.06	&	7.63E-5	&1,1,2,2	\\
050904 (2)	&	6.29	&	857	&	284	&	6.07E+48	&	0.045	&	30.9	&	244.4	&	0.32	&	4.78E-5	&	0.10	&	1.07E-5	&1,1,2,2	\\
050904 (3)	&	6.29	&	1149	&	194	&	3.62E+48	&	0.045	&	24.7	&	214.7	&	0.35	&	2.38E-5	&	0.11	&	4.65E-6	&1,1,2,2	\\
050904 (4)	&	6.29	&	5085	&	3916	&	1.44E+48	&	0.045	&	25.5	&	170.4	&	0.34	&	1.86E-4	&	0.13	&	4.25E-5	&1,1,2,2	\\
050904 (5)	&	6.29	&	16153	&	8713	&	6.96E+47	&	0.045	&	55.9	&	142.2	&	0.25	&	1.01E-4	&	0.15	&	5.11E-5	&1,1,2,2	\\
050904 (6)	&	6.29	&	18383	&	20230	&	1.89E+47	&	0.045	&	34.9	&	102.7	&	0.30	&	9.52E-5	&	0.18	&	4.01E-5	&1,1,2,2	\\
050904 (7)	&	6.29	&	25618	&	5360	&	4.77E+47	&	0.045	&	14.3	&	129.3	&	0.42	&	1.44E-4	&	0.16	&	2.29E-5	&1,1,2,2	\\
051016B	&	0.94	&	374	&	1566	&	2.74E+46	&	0.162	&	22.5	&	63.3	&	0.09	&	1.19E-5	&	0.04	&	8.34E-6	&1,1,2,2	\\
060115	&	3.53	&	369.7	&	83.4	&	5.15E+48	&	0.044	&	41.8	&	234.5	&	0.29	&	1.46E-5	&	0.11	&	4.27E-6	&3,3,2,2	\\
060124 (1)	&	2.3	&	283	&	361	&	9.63E+49	&	0.071	&	6.9	&	487.6	&	0.40	&	0.00907	&	0.03	&	7.07E-4	&1,1,2,2	\\
060124 (2)	&	2.3	&	644	&	363	&	4.38E+49	&	0.071	&	35.2	&	400.3	&	0.19	&	0.00103	&	0.03	&	3.35E-4	&1,1,2,2	\\
060210 (1)	&	3.91	&	136.5	&	60.7	&	1.13E+50	&	0.028	&	45.9	&	507.1	&	0.39	&	1.81E-4	&	0.12	&	2.74E-5	&3,3,2,2	\\
060210 (2)	&	3.91	&	350.9	&	60.2	&	5.93E+49	&	0.028	&	63.3	&	431.9	&	0.35	&	7.01E-5	&	0.13	&	1.57E-5	&3,3,2,2	\\
060418	&	1.49	&	121.9	&	25	&	1.08E+50	&	0.025	&	75.6	&	501.9	&	0.36	&	8.72E-5	&	0.14	&	1.95E-5	&3,3,2,2	\\
060526 (1)	&	3.22	&	96.2	&	24.5	&	2.84E+50	&	0.085	&	12.3	&	639.1	&	0.27	&	8.76E-4	&	0.01	&	1.46E-4	&3,3,2,2	\\
060526 (2)	&	3.22	&	257.5	&	37.8	&	1.44E+50	&	0.085	&	20	&	539.1	&	0.22	&	4.61E-4	&	0.02	&	1.16E-4	&3,3,2,2	\\
060526 (3)	&	3.22	&	279.5	&	49.9	&	1.18E+50	&	0.085	&	17.9	&	512.7	&	0.23	&	5.44E-4	&	0.02	&	1.26E-4 &3,3,2,2		\\
060526 (4)	&	3.22	&	316.8	&	71.9	&	3.93E+49	&	0.085	&	29.1	&	389.8	&	0.18	&	1.81E-4	&	0.02	&	6.3E-5	&3,3,2,2	\\
060707	&	3.425	&	174.5	&	34.6	&	4.87E+48	&	0.149	&	12.2	&	231.3	&	0.15	&	2.67E-5	&	0.01	&	1.02E-5	&3,3,2,2	\\
060714 (1)	&	2.711	&	103.9	&	49.6	&	1.09E+50	&	0.027	&	84.2	&	502.7	&	0.32	&	1.08E-4	&	0.13	&	2.78E-5	&3,3,2,2	\\
060714 (2)	&	2.711	&	55.6	&	8.2	&	7.58E+49	&	0.027	&	38.1	&	459.2	&	0.43	&	2.58E-5	&	0.13	&	3.38E-6	&3,3,2,2	\\
060714 (3)	&	2.711	&	131.5	&	14	&	1.11E+50	&	0.027	&	54.9	&	505.2	&	0.38	&	4.57E-5	&	0.12	&	7.98E-6	&3,3,2,2	\\
060714 (4)	&	2.711	&	151.5	&	21.1	&	8.52E+49	&	0.027	&	46.1	&	473	&	0.40	&	6.23E-5	&	0.13	&	9.59E-6	&3,3,2,2	\\
060729	&	0.54	&	163.4	&	43.7	&	1.60E+49	&	0.418	&	4.5	&	311.3	&	0.07	&	0.00142	&	0.00	&	8.72E-4	&3,3,2,2	\\
060814	&	0.84	&	119.5	&	39	&	1.48E+49	&	0.064	&	28.1	&	305.2	&	0.24	&	7.62E-5	&	0.05	&	1.99E-5	&3,3,2,2	\\
060906 (1)	&	3.69	&	908	&	1460	&	9.68E+46	&	0.027	&	66.9	&	86.8	&	0.35	&	2.75E-6	&	0.32	&	2.17E-6 &4,4,2,2		\\
060906 (2)	&	3.69	&	6200	&	5810	&	5.84E+46	&	0.027	&	95.6	&	76.5	&	0.30	&	4.79E-6	&	0.33	&	5.84E-6 &4,4,2,2		\\
070306	&	1.5	&	169.7	&	43.4	&	2.76E+49	&	0.079	&	25.3	&	356.8	&	0.21	&	1.38E-4	&	0.03	&	4.05E-5	&3,3,2,2	\\
070318 (1)	&	0.84	&	179	&	28.9	&	6.43E+47	&	0.163	&	8.8	&	139.4	&	0.17	&	9.33E-6	&	0.02	&	3.36E-6	&3,3,2,2	\\
070318 (2)	&	0.84	&	203.7	&	147.9	&	1.63E+48	&	0.163	&	9.3	&	176	&	0.16	&	1.17E-4	&	0.01	&	4.25E-5	&3,3,2,2	\\
070721 B (1)	&	3.626	&	256.2	&	121.7	&	1.30E+49	&	0.021	&	561.6	&	295.7	&	0.16	&	5.05E-6	&	0.23	&	8.09E-6	&3,3,2,2	\\
070721 B (2)	&	3.626	&	297.8	&	11.7	&	5.87E+49	&	0.021	&	62.4	&	430.8	&	0.43	&	1.39E-5	&	0.19	&	2.64E-6	&3,3,2,2	\\
070721 B (3)	&	3.626	&	328.7	&	27.2	&	1.84E+49	&	0.021	&	69.5	&	322.6	&	0.41	&	9.18E-6	&	0.22	&	2.39E-6	&3,3,2,2	\\
070721 B (4)	&	3.626	&	575	&	242	&	2.40E+48	&	0.021	&	82.4	&	193.7	&	0.39	&	9.08E-6	&	0.28	&	4.21E-6	&3,3,2,2	\\
071031 (1)	&	2.69	&	66.6	&	61.2	&	7.39E+49	&	0.061	&	19.8	&	456.4	&	0.30	&	3.94E-4	&	0.04	&	6.53E-5	&3,3,2,2	\\
071031 (2)	&	2.69	&	188.1	&	44	&	2.94E+49	&	0.061	&	31.8	&	362.4	&	0.24	&	7.57E-5	&	0.05	&	1.98E-5	&3,3,2,2	\\
071031 (3)	&	2.69	&	242.8	&	51.9	&	1.68E+49	&	0.061	&	34.1	&	315.2	&	0.23	&	4.82E-5	&	0.05	&	1.39E-5 &3,3,2,2		\\
071031 (4)	&	2.69	&	352.7	&	276.1	&	1.11E+49	&	0.061	&	29.5	&	284.3	&	0.25	&	1.91E-4	&	0.06	&	5.04E-5	&3,3,2,2	\\
080210	&	2.641	&	166.6	&	88.5	&	1.03E+49	&	0.031	&	51.1	&	278.8	&	0.35	&	2.99E-5	&	0.15	&	7.65E-6	&5,5,6,6	\\
080810 (1)	&	3.35	&	80.2	&	53	&	7.15E+49	&	0.021	&	49.3	&	452.8	&	0.46	&	1.02E-4	&	0.18	&	1.49E-5	&5,5,6,6	\\
080810 (2)	&	3.35	&	198.2	&	49.7	&	2.19E+49	&	0.021	&	91	&	336.8	&	0.38	&	1.65E-5	&	0.21	&	5.35E-6	&5,5,6,6	\\
080928 (1)	&	1.692	&	148.7	&	201	&	3.21E+49	&	0.038	&	40.2	&	370.7	&	0.33	&	3.69E-4	&	0.10	&	7.42E-5	&5,5,6,6	\\
080928 (2)	&	1.692	&	326	&	80.5	&	7.38E+48	&	0.038	&	33.6	&	256.5	&	0.35	&	4E-5	&	0.12	&	8.3E-6	&5,5,6,6	\\
081008	&	1.9685	&	279.9	&	140.2	&	6.98E+48	&	0.038	&	66.3	&	252.9	&	0.27	&	3.29E-5	&	0.13	&	1.25E-5 &5,5,6,6	\\
100906A	&	1.727	&	76.6	&	130.6	&	1.72E+50	&	0.029	&	49.6	&	563.7	&	0.37	&	9.99E-4	&	0.10	&	1.58E-4	&5,5,6,6	\\
110801A (1)	&	1.858	&	192.3	&	51.9	&	6.24E+48	&	0.035	&	33.5	&	246	&	0.37	&	2.03E-5	&	0.14	&	4.09E-6	&5,5,6,6	\\
110801A (2)	&	1.858	&	317.2	&	307.6	&	3.12E+49	&	0.035	&	71.9	&	368	&	0.28	&	3.05E-4	&	0.11	&	9.61E-5	&5,5,6,6	\\
121024A (1)	&	2.298	&	181.9	&	69.5	&	4.35E+48	&	0.059	&	20.7	&	224.7	&	0.30	&	2.81E-5	&	0.07	&	5.79E-6 &5,5,6,6		\\
121024A (2)	&	2.298	&	254.6	&	69.4	&	2.67E+48	&	0.059	&	27.9	&	198.8	&	0.26	&	1.33E-5	&	0.08	&	3.73E-6	&5,5,6,6	\\
121211A	&	1.023	&	116.7	&	748	&	3.93E+48	&	0.078	&	47.7	&	219.1	&	0.15	&	2.67E-4	&	0.05	&	1.34E-4	&5,5,6,6	\\
130427B	&	2.78	&	115.3	&	54.5	&	8.00E+48	&	0.038	&	45.2	&	261.7	&	0.31	&	1.6E-5	&	0.12	&	4.29E-6	&5,5,6,6	\\
130606A (1)	&	5.91	&	83.6	&	13.2	&	7.73E+49	&	0.022	&	42.7	&	461.8	&	0.47	&	2E-5	&	0.17	&	2.57E-6	&5,5,6,6	\\
130606A (2)	&	5.91	&	73	&	108.9	&	7.53E+49	&	0.022	&	22.2	&	458.5	&	0.56	&	3.03E-4	&	0.17	&	2.07E-5	&5,5,6,6	\\
130606A (3)	&	5.91	&	196.7	&	56.4	&	6.68E+49	&	0.022	&	50.9	&	445	&	0.44	&	6.24E-5	&	0.17	&	9.72E-6	&5,5,6,6	\\
130606A (4)	&	5.91	&	240.7	&	143.2	&	6.73E+49	&	0.022	&	120.8	&	446	&	0.33	&	7.16E-5	&	0.17	&	2.48E-5	&5,5,6,6	\\
130606A (5)	&	5.91	&	347	&	125.1	&	4.12E+49	&	0.022	&	44.8	&	394.3	&	0.46	&	9.65E-5	&	0.19	&	1.45E-5	&5,5,6,6	\\
140512A 	&	0.725	&	99.5	&	76.9	&	1.05E+49	&	0.044	&	29.6	&	279.9	&	0.33	&	9.79E-5	&	0.10	&	1.93E-5	&5,5,6,6	\\
\hline
\end{tabular}

\footnotesize
$^a$ In units of seconds.\\
$^b$ In units of erg/s.\\
$^c$ In units of rad.\\
$^d$ In units of $M_{\odot}$.\\
$^e$ References for $T_{\rm start}$, duration, $\theta_{\rm j}$, and the Lorentz factor of X-ray flares.

References - 1. \cite{2007ApJ...671.1921F}; 2. \cite{2015ApJ...807...92Y}; 3. \cite{2010MNRAS.406.2113C}; 4. \cite{2011A&A...526A..27B}; 5. \cite{2016ApJS..224...20Y}; 6. \cite{2017RAA....17...53X}.
\end{table*}
%%%%%%%%%%%%%%%%% APPENDICES %%%%%%%%%%%%%%%%%%%%%

%%%%%%%%%%%%%%%%%%%%%%%%%%%%%%%%%%%%%%%%%%%%%%%%%%

\end{document}